\documentclass[fleqn,twoside]{article}
\usepackage{amsmath}
\usepackage[headings]{espcrc2}

\readRCS
$Id: espcrc2.tex,v 1.2 2004/02/24 11:22:11 spepping Exp $
\usepackage{graphicx}
\usepackage[figuresright]{rotating}

\title{The effective action and the triple Pomeron vertex  }

\author{Martin Hentschinski \address[MCSD]{
Instituto de F\'isica Te\'orica UAM/CSIC, \\
Universidad Aut\'onoma de Madrid, 28049 Madrid, Spain \\
} \address[MSCD]{
II. Institute for Theoretical Physics, Hamburg University,  \\ 
Luruper Chaussee 149, 22761 Hamburg, Germany        
}%
        \thanks{This work has been carried out in collaboration with J. Bartels and L.N. Lipatov. Support by the Graduiertenkolleg ``Zuk\"unftige Entwicklungen in der Teilchenphysics'' and DESY is acknowlegded.}}

\runtitle{The effective action and the triple Pomeron vertex}
\runauthor{M. Hentschinski}

\begin{document}

\begin{abstract}
We study integrations over light-cone momenta in the gauge invariant effective action of high energy QCD. A regularization mechanism which allows for the evaluation of the longitudinal integrations is presented. After a rederivation of the reggeized gluon and the BFKL-equation from the effective action, we study the $1-3$ and $2-4$ reggeized gluon transition vertex of QCD Reggeon field theory  and discuss their connection with the usual triple Pomeron vertex of perturbative QCD.
\vspace{1pc}
\end{abstract}

\maketitle

\section{Introduction}
\label{sec:simple}

In 1995 an effective action \cite{action} for QCD scattering processes
at high center of mass energies $\sqrt{s}$ has been proposed by L.N.
Lipatov which describes the interaction of fields of reggeized gluons
($A_\pm = -it^aA^a_\pm$ ) with quark ($\psi$) and gluon ($v_\mu =-it^a
v^a_\mu$) fields, local in rapidity. The effective action reads
  \begin{eqnarray}
  \label{eq:effact1}
  {S}_{\mbox{eff}} 
= \int d^4 x \big(
  \mathcal{L}_{\text{QCD}} (v_\mu, \psi ) 
+  
 \mathcal{L}_{\text{ind}} (v_\pm, A_\pm )\big),
\end{eqnarray}
where $\mathcal{L}_{\text{QCD}}$ is the usual QCD-Lagrangian  and
 \begin{eqnarray}
\label{eq:induced}
\mathcal{L}_{\text{ind}}(v_\pm, A_\pm )  &=& \mbox{tr}[(A_-(v)\! -\!A_-)\partial^2 A_+ ]
\nonumber \\
&& + \mbox{tr}[ (A_+(v)\! -\! A_+)\partial^2 A_-]\big)
\end{eqnarray}
is the {\it induced} term with 
\begin{align}
\label{eq:induced2}
  A_\pm(v)  &= v_\pm D_\pm^{-1}\partial_\pm =
\notag  \\ 
  =v_\pm &- gv_\pm\frac{1}{\partial_\pm}v_\pm + g^2v_\pm\frac{1}{\partial_\pm}v_\pm\frac{1}{\partial_\pm}v_\pm - \ldots
\end{align}
Light-cone components are defined by $k^\pm \equiv n^\pm\cdot k$ where
$n^{\pm}$ are the light cone directions associated with the scattering
particles.  The reggeized gluon fields $ A_\pm$ are bare reggeized
gluons with trajectory $j(t) = 1$, while reggeization occurs as a
higher loop correction. The fields $A_{\pm}$ have the special property to be
invariant under local gauge transformations, even though they
transform globally in the adjoint representation of $SU(N_c)$. The
effective action allows then to factorize high energy QCD amplitudes
into gauge invariant pieces which themselves are localized in
rapidity. In particular, the interaction between particles and
reggeized gluon fields, is by definition restricted to a small
rapidity interval $\Delta Y < \eta$, while all non-local interaction,
which extends over rapidity intervals $\Delta Y > \eta$, is mediated
by  reggeized gluons.

Due to these properties the effective action promises to be a suitable
tool to study both higher order corrections to the BFKL-equation and
 unitarization of the BFKL-Pomeron.  However, due the
factorization of high energy QCD amplitudes into pieces local in
rapidity, $ Y = \ln(k^+/k^-)/2$, loop integrals over (longitudinal)
light-cone momenta take a special role in the effective action. In
particular, longitudinal integrals require additional regularization as
otherwise the locality principle of the effective action is violated.
Transverse integrals on the other hand may be treated by conventional
methods.  Furthermore, there exists sometimes the danger of
overcounting certain parts of the underlying QCD amplitude within the
effective action which can then lead to the occurrence of divergent
integrals. In Sec.\ref{sec:bfkl} we demonstrate how these points can
be addressed in the context of quark-quark scattering within the
Leading Logarithmic Approximation (LLA). In Sec.\ref{sec:rft} we apply
the obtained rules to the derivation of reggeized gluon transition
vertices and compare them with previous results.

\section{The reggeized gluon and the BFKL-Pomeron in the effective action}
\label{sec:bfkl}

\begin{figure}[t]
\parbox{2cm}{\center \includegraphics[height=2.3cm]{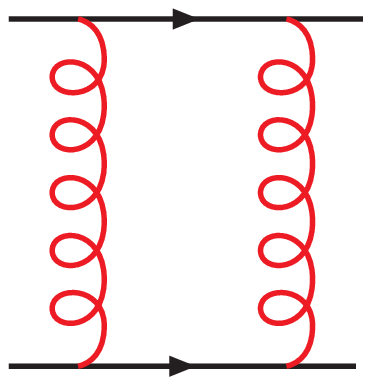}}
 \parbox{2cm}{\center \includegraphics[height=2.3cm]{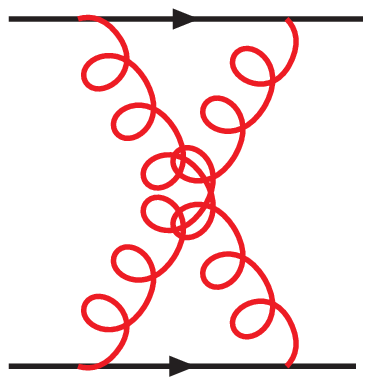}}
\parbox{.5cm}{ $\,$}
\parbox{2cm} {\center \includegraphics[height=2.3cm]{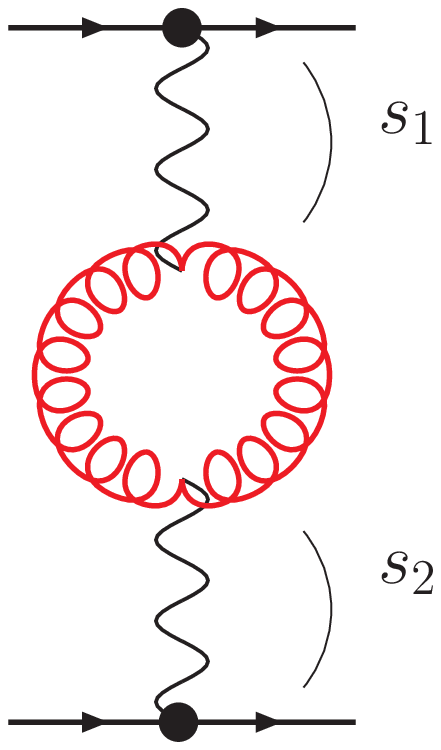}}
\\
\parbox{4.2cm}{\center (a)} \parbox{.5cm}{ $\,$} \parbox{2cm}{ \center (b)}
 \caption{\small (a) QCD 1-loop diagrams that yield the leading logarithmic contribution. Their leading ${\bf 8_A}$-sector  is in the effective action shifted to the loop-correction of the reggeized gluon (b).}
  \label{fig:redistr}
\end{figure}
\begin{figure}[t]
  \centering
  \parbox{1.7cm}{ \includegraphics[width=1.7cm]{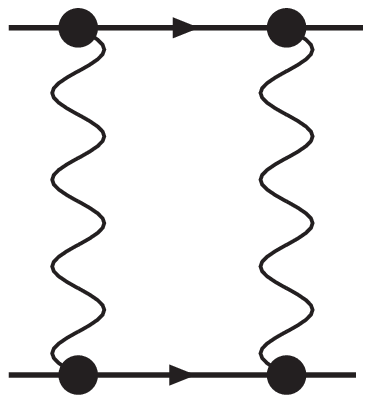}}
\parbox{1.7cm}{ \includegraphics[width=1.7cm]{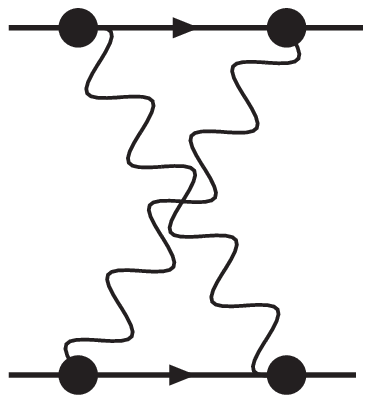}}
\parbox{1.7cm}{ \includegraphics[width=1.7cm]{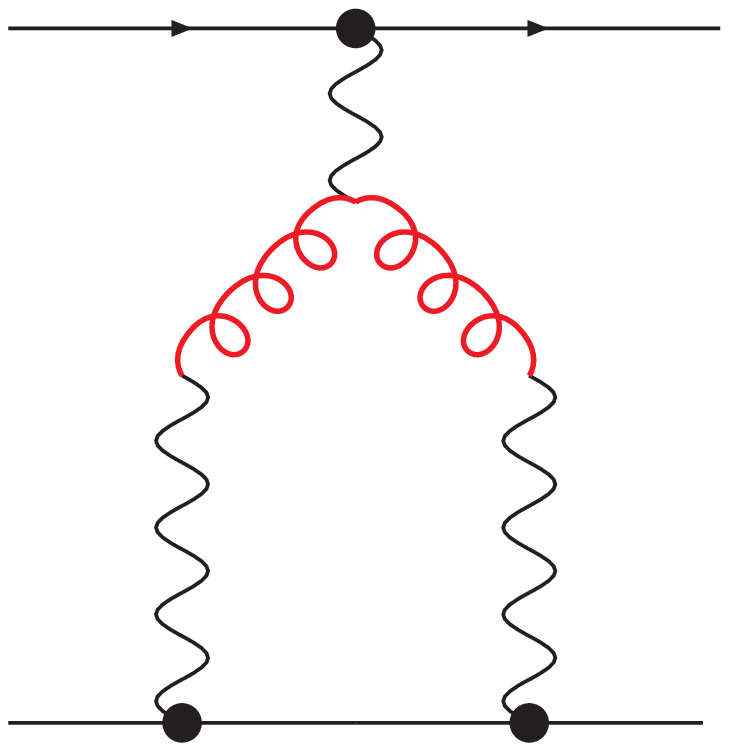}}
 \parbox{1.7cm}{\includegraphics[width=1.7cm]{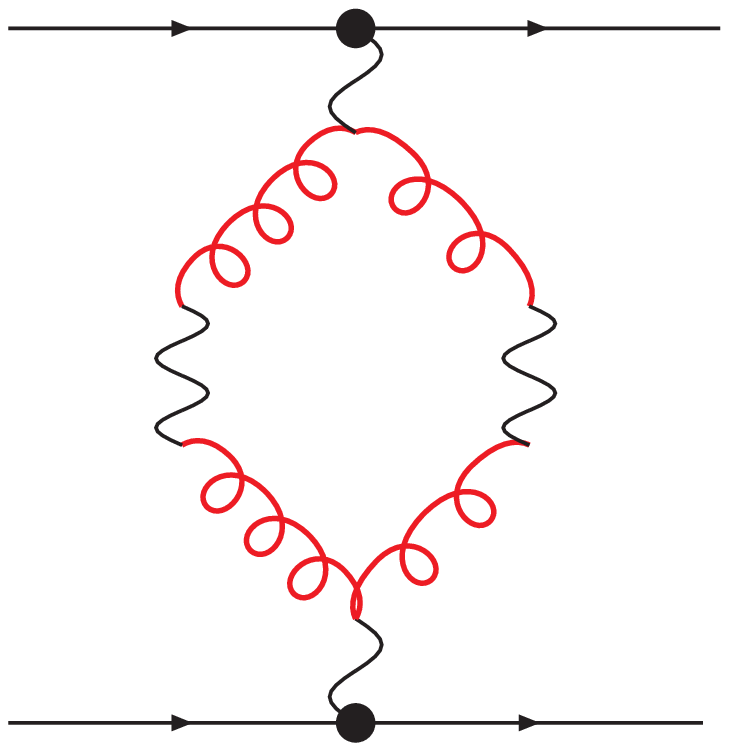}}
  \caption{\small Typical diagrams  with exchange of two reggeized gluons in the effective action.  }
  \label{fig:tworegg}
\end{figure}

The Lagrangian of the effective action, Eq.~(\ref{eq:effact1}),
consists apart from the usual QCD Lagrangian of the induced term,
Eq.~(\ref{eq:induced}). This additional term leads in the effective
action to a redistribution of the antisymmetric color octet sector,
${\bf 8}_A$, of the underlying QCD Feynman diagrams.  A 1-loop example
for such a re-distribution is shown in Fig.~\ref{fig:redistr}.  There, in the effective theory diagram Fig.~\ref{fig:redistr}.b,
the coupling of the reggeized gluon to the gluon loop takes place by
the induced vertex $-gv_\pm\frac{1}{\partial_\pm}v_\pm \partial^2
A_\mp$ which arises from Eq.~(\ref{eq:induced2}).  The diagram
Fig.~\ref{fig:redistr}.b is of particular importance in the effective
action, as it yields the 1-loop correction to the trajectory of the
reggeized gluon.  In order to evaluate the resulting (longitudinal)
loop-integrations of Fig.~\ref{fig:redistr}.b, it is necessary to take
into account that the exchange  of the reggeized gluons demands the
squared center-of-mass energies $s_1$ and $s_2$ of the regarding
sub-amplitudes to be large, {\it i.e.} a corresponding lower bounds needs to be imposed.  A simple lower cut-off  however
misses the imaginary part of Fig.~\ref{fig:redistr}.b.  It is
therefore necessary  to use a more elaborated method of regularization.
A suitable choice turns out to be the following Mellin-integral
\begin{align}
  \label{eq:mellin}
 \int_{0-i\infty}^{0+i\infty} \frac{d\omega}{4\pi i}& \frac{1}{\omega + \nu} \left[ \left(\frac{- s_1 - i\epsilon}{\Lambda}\right)^\omega + \left(\frac{ s_1 - i\epsilon}{\Lambda}\right)^\omega \right]& \nonumber \\
&= \phantom{\bigg(}\theta(|s_1/\Lambda - 1| )& \qquad \qquad \,
\end{align}
where we take the parameter $\nu > 0$ in the limit $\nu \to 0$.  In
the above expression, $-\nu$ has the interpretation of an
infinitesimal small Regge trajectory, while the integration variable
$\omega$ takes the role of complex angular momentum. This kind of
regularization has the great advantage that it allows to take into
account in a rather straight forward way phases in $s_1$ (and
therefore also imaginary parts). In particular, negative signature of
the reggeized gluon is made explicit.  Making use of
Eq.~(\ref{eq:mellin}), the diagram Fig.~\ref{fig:redistr}.b can be
evaluated and we obtain within the LLA the well-known result for the
reggeized gluon at 1-loop:
\begin{eqnarray}
  \label{eq:traj}
  {\cal M}_{\text{Fig.}\ref{fig:redistr}.b}(s,t) ={\cal M}_{\text{tree}}(s,t) \beta(t) \frac{\ln (-s) + \ln s}{2}.
\end{eqnarray}
Here $ {\cal M}_{\text{tree}}$ is the quark-quark scattering amplitude
at tree-level and $\beta(t)$ is the well-known 1-loop correction to
the gluon trajectory. Resummation of diagrams like
Fig.~\ref{fig:redistr}.b, with an arbitrary number of gluon loops
inserted, yields then within the LLA the all order reggeized gluon
with negative signature.  The leading contribution of the $t$-channel
exchange with positive signature arises on the other hand from the
exchange of two reggeized gluons.  Typical diagrams with two reggeized
gluon exchange that arise from the effective action at 1-loop are
shown in Fig.~\ref{fig:tworegg}.
Taking a closer look on the resulting expressions and comparing them
with the underlying QCD graphs, Fig.~\ref{fig:redistr}.a, it turns out
that they all contain a term which occurs already in the diagram
Fig.~\ref{fig:redistr}.b. Moreover, due to the simplified structure of
the reggeized gluon propagator in the diagrams Fig.~\ref{fig:tworegg},
this term leads for every individual diagram to a divergence in the
longitudinal part of the loop integral. While in the case of two
reggeized gluon exchange this problem can be cured by a principal
value prescription (see for instance \cite{talks}), such a procedure
turns out to fail if states with more than two reggeized gluon are
considered. A suitable way to deal with this problem is therefore to
remove the overcounted terms completely from the regarding diagrams.
For general diagrams that contain the exchange of $n$ reggeized
gluons, this can be achieved by supplementing the Lagrangian of the
effective action by a term
\begin{align}
  \label{eq:supp}
{\cal L}_{\text{supp}} (A_+, A_-) = -2 {\cal L}_{\text{ind}} (A_\pm, A_\pm).
\end{align}
Note that adding such a term is not in conflict with the original
derivation of the effective action. With this term we obtain for the
exchange of two reggeized gluons the following result
\begin{align}
  \label{eq:tworegg_wloop}
{\cal M} &= 2\pi i |s| A^{a_1a_2}_{(2,0)} \otimes_{12}  A^{a_1a_2}_{(2,0)}, 
\end{align}
 which carries explicitly positive signature. There, the two (reggeized) gluon  impact factor $ A_{(2,0)},$ is  given by
\begin{align}
  \label{eq:impa2}
 A^{a_1a_2}_{(2,0)} = -g^2 \frac{1}{2} \left( \frac{1}{N_c} \delta^{a_1a_2} + d^{a_1a_2c} t^c_{AA'} \right),
\end{align}
while $ \otimes_{12} = \int \frac{d^2 {\bf k}}{(2\pi)^3 {\bf k}_1^2
  {\bf k}_2^2} $.  The state of two reggeized gluons couples therefore
in the color singlet, ${\bf 1}$, and the symmetric color octet, ${\bf
  8_S}$, to the quark, while the ${\bf 8_A}$-sector decouples and is
contained in the single reggeized gluon exchange.  
Higher order corrections within the LLA  include
apart from loop corrections to the individual  reggeized gluons,
Eq.~(\ref{eq:traj}), also the interaction between the reggeized gluons, which yields the (real part) of the BFKL-kernel.
  Resumming both types of corrections
within the LLA, the famous BFKL-equation is re-obtained. The color
singlet sector then yields the BFKL-Pomeron, while in the ${\bf 8_S}$
sector further reggeization occurs.  Unlike the ${\bf 8_A}$-reggeized
gluon, the ${\bf 8_S}$-Reggeon is not a fundamental degree of freedom
in the effective action, but arises as a state of two reggeized
gluons.

\section{Vertices in QCD Reggeon field theory}
\label{sec:rft}
\begin{figure}[t]
  \centering
 \parbox{2.9cm}{ \includegraphics[height=1.8cm]{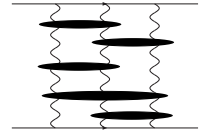} }
  \caption{\small  BKP-state of three reggeized gluons with the number of reggeized gluons in the $t$-channel conserved. }
  \label{fig:bfkl}
\end{figure}
\begin{figure}[t]
  \centering
  \parbox{3cm}{\center {\Large $\sum$}
    \parbox{2cm}{\includegraphics[height=1.2cm]{a4_1loop.eps}}} 
  \parbox{3cm}{\center \includegraphics[height=1.3cm]{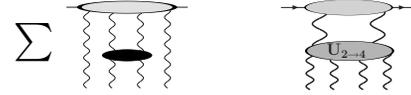}}
  \caption{\small The state of four reggeized gluons in the $t$-channel may either arise from four reggeized gluons which couple directly to the quark and then interacted pairwise by the BFKL-Kernel  and from a two-to-four transition.}
  \label{fig:fourstate}
\end{figure}

Beyond the state of two reggeized gluons, one is at first lead to
so-called BKP-states {\it i.e}. states of $n$ reggeized gluons, with
the number of reggeized gluons in the $t$-channel conserved. Within the LLA the
reggeized gluons interact pairwise by the BFKL-Kernel, see
Fig.~\ref{fig:bfkl}, and the whole system is known  to be integrable in
the large $N_c$ limit.
 In a next step one
should further take into account transition vertices which change the number of
the reggeized gluons in the $t$-channel.  With the
rules for longitudinal integrations introduced above, the effective
action can  be used to derive these  vertices.
\begin{figure}[t]
  \centering
   \parbox{2.7cm} {\includegraphics[height=1.6cm]{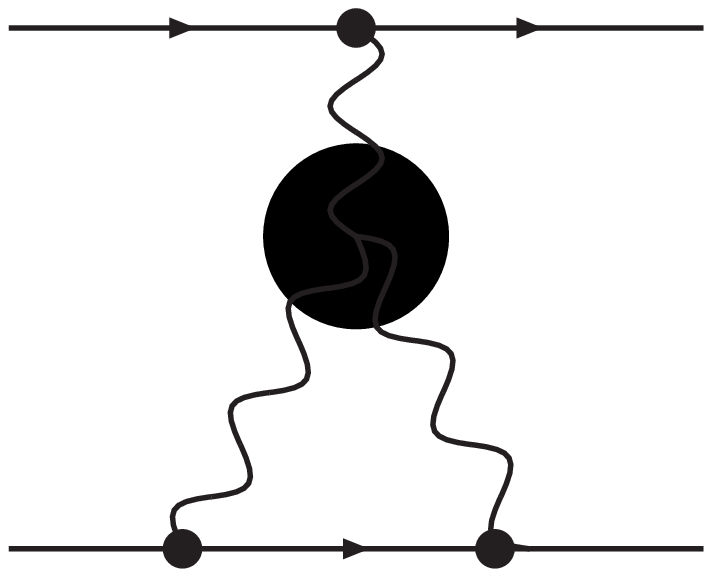}}
 \parbox{2.7cm} {\includegraphics[height=1.6cm]{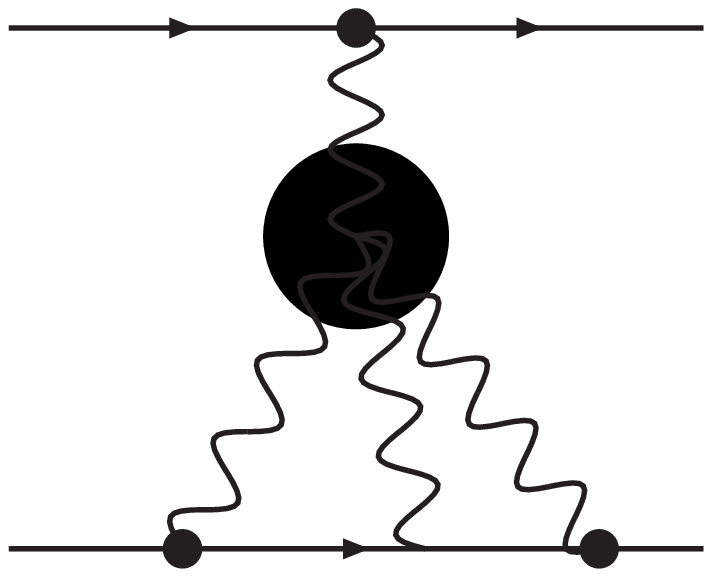}}
  \caption{\small Transition of  1 to 2 and 1 to 3 reggeized gluons inside the elastic quark-quark scattering amplitude. }
  \label{fig:123}
\end{figure}
As pointed out by Gribov in \cite{gribov}, signature is conserved
within the elastic scattering amplitude. Generally, odd number state
of reggeized gluons carry negative signature and even number states of
reggeized gluons positive signature. The transition of one to two
reggeized gluons, Fig.~\ref{fig:123}, contradicts therefore signature
conservation. At 1-loop this requirement turns out to be automatically
fulfilled in the effective action, as the one-to-two transition
vanishes inside the elastic scattering amplitude. The transition from
one-to-three reggeized gluons is on the other hand allowed by
signature conservation and a non-zero transition vertex $U_{1 \to 3}$
can be derived from the effective action. A similar result holds for
the transition from two-to-three an two-to-four reggeized gluons.
While the former vanishes if inserted in the elastic scattering
amplitude, the latter is allowed by signature conservation and yields
a non-zero vertex\footnote{For explicit results we refer the reader to
  \cite{thesis}}, $U_{2 \to 4}$. The transition vertex $U_{2 \to 4}$
has however no good infrared properties, not even if restricted to the
overall color singlet. Good IR-behavior is only obtained if one takes
into account the complete set of diagrams that yield a state of four
reggeized gluons in the $t$-channel, Fig.~\ref{fig:fourstate}.  The
four reggeized gluon state coupling directly to the quark can be shown
to yield (in the overall color singlet) further reggeization in the
symmetric color sector plus a term which takes the form of another
two-to-four transition:
\begin{align}
   { \sum} \parbox{1.5cm}{ \includegraphics[height=1.2cm]{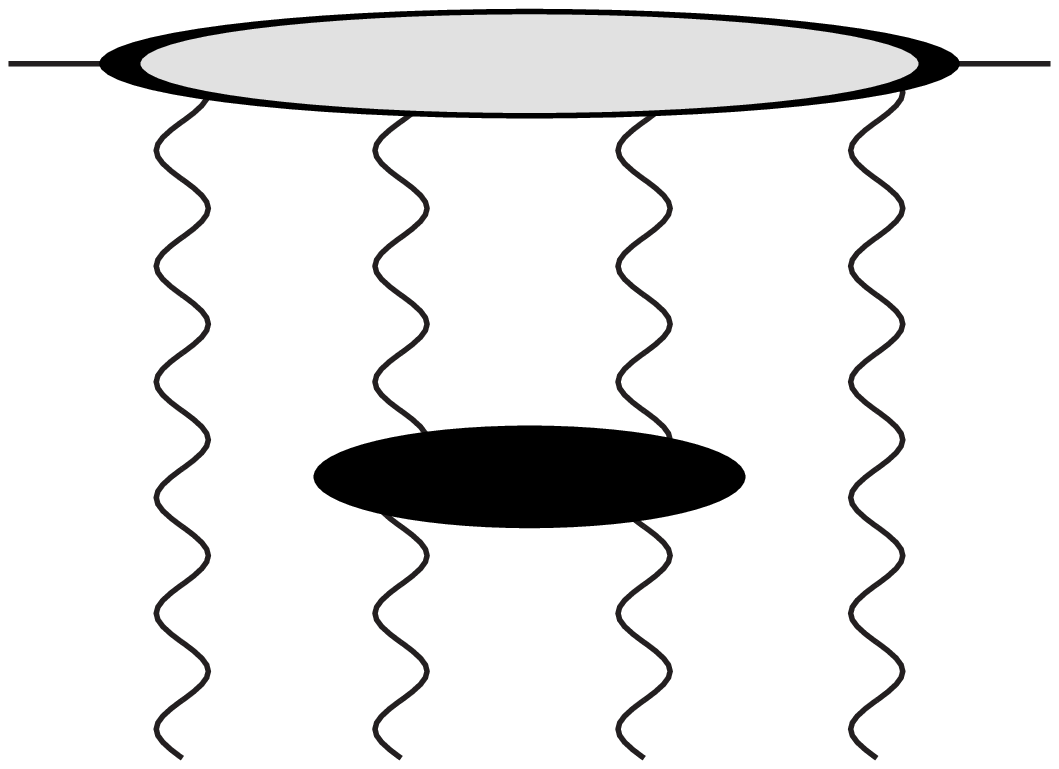}} &= 
\parbox{1.65cm}{ \includegraphics[height=1.3cm]{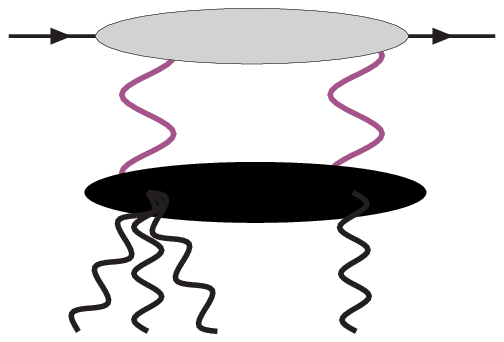}}
+  \parbox{1.65cm}{ \includegraphics[height=1.3cm]{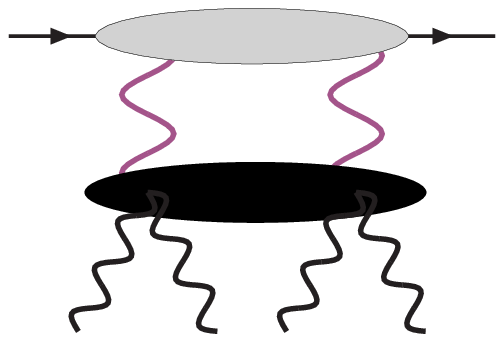}}
\notag \\ 
&+ \text{perm.} + 
\parbox{1.65cm}{\includegraphics[height=1.3cm]{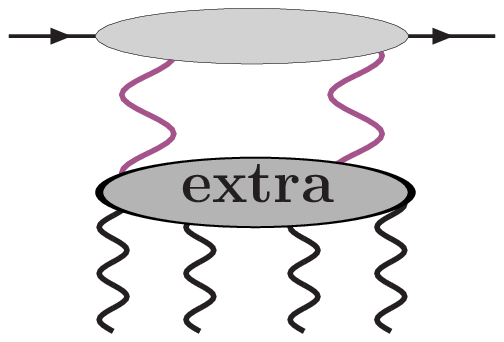}
}
\end{align}
Combining  this extra term with the two-to-four reggeized gluon vertex $U_{2 \to 4}$, one obtains the vertex $V_{2 \to 4}$:
\begin{align}
  \parbox{1.75cm}{\includegraphics[height=1.3cm]{u24_loop.ps}}
 +
 \parbox{1.75cm}{\includegraphics[height=1.3cm]{extra_loop_alt.eps}} = 
\parbox{1.75cm}{\includegraphics[height=1.3cm]{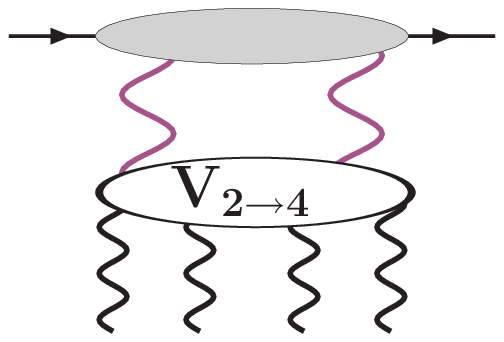}}
\end{align}
This vertex $V_{2 \to 4}$ is well known from the study of the triple
discontinuity of a six-point amplitude in \cite{wuest} and is in
particular infrared finite. Projecting pairs of reggeized gluons on
the color singlet, $V_{2 \to 4}$ yields then the triple Pomeron
vertex.

In this contribution we presented a set of rules which allows to carry
out longitudinal integration in the effective action,
Eq.~(\ref{eq:effact1}). These rules have been successfully applied to
the derivation of reggeized gluon transition vertices from the
effective action within the LLA and the obtained expressions are in agreement
with earlier results. Future research should concentrate on extending
these rules beyond the LLA.

\end{document}